\documentclass[aps,prl,superscriptaddress,notitlepage,nopacs,amsmath,amstex,amssymb,citeautoscript,longbibliography,floatfix,letter,twocolumn]{revtex4-2}
\usepackage{graphicx}
\usepackage{subfigure}
\usepackage{adjustbox}
\usepackage{bm}
\usepackage{color}
\usepackage{braket}
\usepackage{standalone}
\usepackage{multirow}
\usepackage{tikz}
\usepackage{mathrsfs}
\usepackage{dsfont}
\usepackage{comment}
\usepackage{amsmath}
\usepackage{ulem}
\usepackage[colorlinks,bookmarks=true,citecolor=blue,linkcolor=red,urlcolor=blue]{hyperref}
\usepackage{cleveref}
\usepackage{orcidlink}

\usepackage{diagbox}
\usepackage{lipsum} 
\AtBeginDocument{\pagenumbering{arabic}}

\begin{document}

\title{Non-Abelian fusion and braiding in many-body parton states}

\author{Koyena Bose}
\email{koyenab@imsc.res.in}
\affiliation{Institute of Mathematical Sciences, CIT Campus, Chennai, 600113, India}
\affiliation{Homi Bhabha National Institute, Training School Complex, Anushaktinagar, Mumbai 400094, India}

\date{\today}

\begin{abstract}

Fractional quantum Hall (FQH) states host fractionally charged anyons with exotic exchange statistics. Of particular interest are FQH phases supporting non-Abelian anyons, which can encode topologically protected quantum information. In this work, we construct quasihole bases for a broad family of non-Abelian FQH states using parton wave functions, which reproduces the fusion-space dimensionality expected from their underlying conformal field theory, consistent with level–rank duality across the parton family. As an application, we numerically compute braiding matrices for representative parton states for large systems, providing a general framework for diagnosing non-Abelian characteristics in candidate FQH states.

\end{abstract}	
\maketitle

\textbf{\textit{Introduction.}} When electrons confined to two dimensions are subjected to a strong magnetic field, their kinetic energy is quenched into discrete Landau levels (LLs), such that Coulomb interactions set the dominant energy scale. In this regime, interactions lift the extensive LL degeneracy and select a unique correlated ground state. At certain fractional LL fillings, this produces an incompressible phase with a finite excitation gap, manifesting as the fractional quantum Hall (FQH) effect~\cite{Tsui82}. The strong correlations in FQH states give rise to numerous interesting properties. Most notably, FQH excitations are fractionally charged objects called anyons~\cite{Laughlin83, de-Picciotto97}, that obey fractional exchange statistics~\cite{Wilczek82, Arovas84, Halperin84, Nakamura20, Bartolomei20, Nakamura23, kim2024aharonovbohm, Samuelson24, Werkmeister24}. At certain filling fractions, these anyons are believed to exhibit non-Abelian statistics, making them promising candidates for fault-tolerant topological quantum computation~\cite{Freedman03, Nayak08}. 

A prominent example is $\nu{=}5/2$~\cite{Willett87}, which is widely believed to host a non-Abelian phase~\cite{Banerjee17b} described by the Moore–Read Pfaffian (MR) state~\cite{Moore91}. Physically, the MR can be viewed as a topological $p$-wave paired state of electron-vortex composites, called composite fermions~\cite{Jain89}. The idea of pairing was generalized to clustering to construct a family of $k$-clustered Read-Rezayi (RR$k$) states [where $k{=}1$ and $k{=}2$ correspond to Laughlin~\cite{Laughlin83} and MR states, respectively], whose wave functions can be formulated as the correlators in $\mathcal{Z}_k$ parafermion conformal field theory (CFT)~\cite{Zamolodchikov85, Gepner87}. The $k{=}3$ Read-Rezayi state is of particular interest due to its potential for universal topological quantum computation, in contrast to the $k{=}2$ state, which does not support a complete set of unitary transformations required for universal quantum gates~\cite{Freedman02, Freedman01}. Encouragingly, the RR$3$ state may be realized experimentally in the $12/5$ FQH state \cite{Read99, Xia04, Rezayi09, Kumar10, Pakrouski16}. For $k{>}2$, the RR$k$ wave functions are difficult to evaluate at large system sizes, but a recent advance has enabled their simulation in this regime~\cite{Bose25a}.

Beyond the RR$k$ sequence, a variety of non-Abelian phases are expected to occur at other filling fractions. Parton theory~\cite{Jain89b} provides a broad class of candidate states, offering a general framework to construct explicit many-body wave functions across a wide range of filling factors in both the lowest and higher LLs~\cite{Wu17, Faugno19, Faugno20a, Faugno21, Balram21, Balram21b, Balram21d, Sharma22, Dora22, Bose23, Sharma23}. Importantly, the parton wave functions are numerically tractable at large system sizes. The possibility of non-Abelian phases within this framework was first established using Chern–Simons (CS) field theory~\cite{Wen91}. In particular, the $\Phi_2^k$ parton state, with $\Phi_2$ denoting two filled LLs, has been shown to have the same anyonic content as RR$k$ states from field theory~\cite{Wen91, Balram19}. While the braiding statistics of MR state with quasiholes (QHs), i.e., positively charged anyons, have been studied microscopically using many-body wave functions~\cite{Nayak96, Ardonne07}, analogous treatments of quasiparticles (QPs) (negatively charged anyons) and of other candidate non-Abelian states remain largely unexplored. In this work, we present a general procedure to construct appropriate QH basis states to demonstrate non-Abelian statistics in parton states. Using this framework, we provide the first many-body wave function-based demonstration of the CS level–rank duality, which states that $\Phi_n^m$ and $\Phi_m^n$ share the same anyonic data. We illustrate the approach for several non-Abelian parton families and numerically confirm the predicted dualities. Finally, we compute the four-QH braiding matrices for the $\Phi_2^2$ and $\Phi_2^3$ states, reaching system sizes up to $N{=}80$ particles. In particular, this allows for the extraction of Ising and Fibonacci-anyon braiding data from numerical simulations of large systems.

\textbf{\textit{Parton Theory.}} In the parton theory put forth by Jain~\cite{Jain89b}, a system of interacting electrons in the FQH regime is described by a collection of $l$ species of non-interacting fractionally charged particles, termed partons. The parton species, labeled by $\lambda{=}1,2,{\cdots},l$, occupies an IQH state at integer filling $n_{\lambda}$, and the resulting many-body wave function for the partonic FQH state, denoted as ``$n_{1}...n_{l}$", is given by~\cite{Jain89b}
\begin{equation} 
\label{eq: parton_wfn} 
\Psi_{\nu}=\mathcal{P}_{\rm LLL} \prod_{\lambda=1}^l \Phi_{n_{\lambda}}(\{ z \}), 
\end{equation} 
where $\Phi_{n_{\lambda}}$ is an integer quantum Hall (IQH) state consisting of $n_\lambda$ filled pseudo Landau levels, called Lambda levels ($\Lambda$Ls), with $\Phi_{{-}n_\lambda}{\equiv}\Phi_{\bar{n}_\lambda}{=}[\Phi_{n_\lambda}]^{*}$. The set of coordinates of the electrons is represented by $\{z\}$, and $\mathcal{P}_{\rm LLL}$ projects the wave function onto the lowest Landau level (LLL) (as is relevant in the strong magnetic field limit). Since the partons experience the same magnetic field as the parent electrons and are at the same density as the electrons, the $\lambda$ parton species carries a fractional charge $q_{\lambda}{=}\nu (-e)/n_{\lambda}$, where $-e$ denotes the elementary charge of the electron. Moreover, since the individual parton charges must sum to that of an electron, the electron and parton fillings are related as $\nu{=}(\sum_{\lambda} n_{\lambda}^{{-}1})^{{-}1}$. 

The possibility for parton states to support non-Abelian anyons was pointed out by Wen~\cite{Wen91}, who showed that a parton state composed of multiple factors of the same IQH state, i.e., having a factor of $\Phi_n^m$, has an effective field theory given by level-$n$ $SU(m)$ CS theory, denoted as $SU(m)_n$. The $SU(m)_n$ CS theory in the bulk is accompanied, via the bulk-boundary correspondence, by a level-$m$ $SU(n)$ Wess–Zumino–Witten (WZW) CFT on the edge, enforced by gauge invariance. A recent study showed how explicit many-body parton states can be generated from the fields of an associated chiral algebra whose fields form the vacuum representation of the $U(1) {\otimes} SU(n)_m$ WZW current algebra~\cite{Henderson24}, thereby building on Wen's results. These field-theoretic predictions can, in principle, be accessed directly at the microscopic level by constructing QH wave functions and extracting the resulting fusion spaces and braiding matrices from their exchange.

\begin{figure}[htbp!]
\includegraphics[clip,width=0.9\columnwidth,height=0.32\columnwidth]{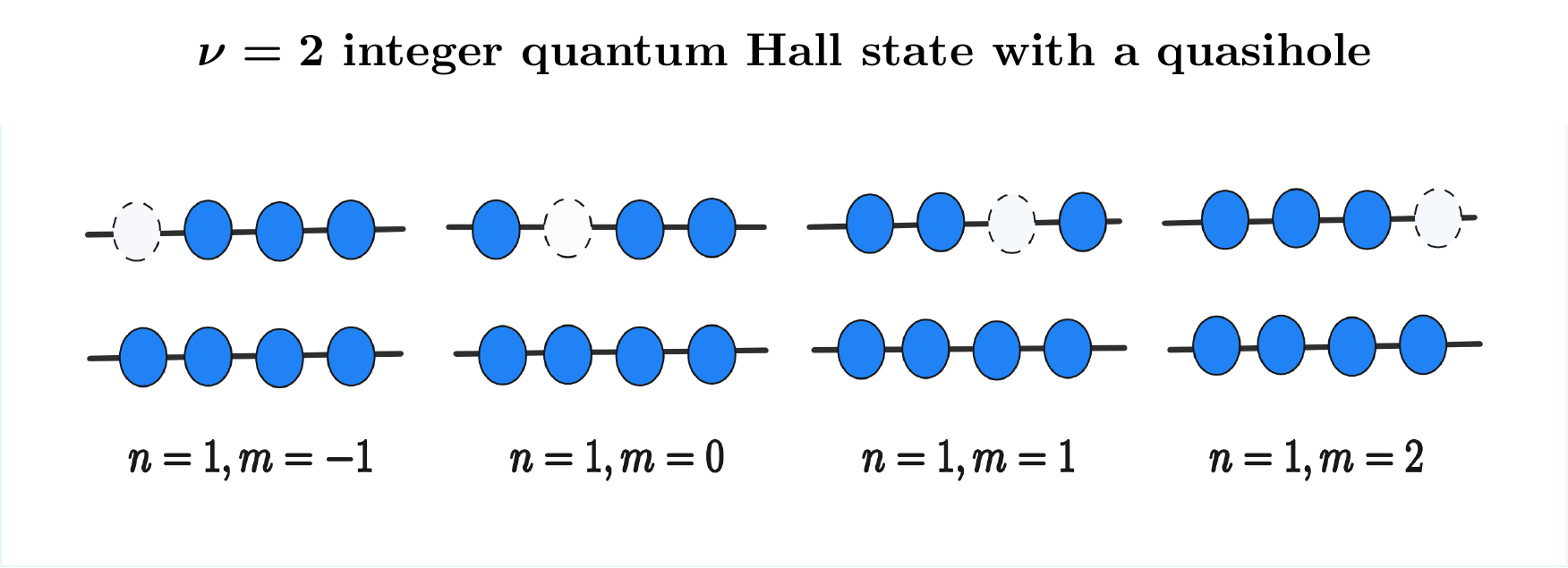} 
\caption{Single-particle occupations for the $\nu{=}2$ integer quantum Hall state for $N{=}8$ electrons, showing different choices of the empty SLL orbital (labeled by $m$).
 }
\label{fig: 2_IQH_with_QH}
\end{figure}

\textbf{\textit{Excitations.}} We now work in the disk geometry, where the electrons are confined to a two-dimensional plane and their positions are parametrized by the complex coordinate $z{=}x{-}iy$. In this geometry, rotational symmetry ensures that the angular momentum about the $z$-axis, denoted by $L_z$, is a conserved quantity. Consequently, single-particle orbitals in $n$th LL, with $n{=}0,1,\ldots$, are labeled by increasing angular momentum $m{=}{-}n,{-}n{+}1,\dots$. A localized QH at $\omega$ in any $n$th IQH factor is given by:
\begin{equation}
\label{eq: QH_coherent_state}
    \Phi^{N_{\rm QHs}{=}1}_{n}=\sum_{m=-(n-1)}^{\infty} (-1)^{m+n-1}  \omega^{m+n-1} \Phi^{N_{\rm QHs}=1,m}_{n} (\{ z \})
\end{equation}
where, $\phi_n^{N_{\rm QHs}=1,m}$ is a Slater determinant of $N$ electrons filling the lowest $n$ LLs except for the $L_{z}{=}m$ orbital in the highest occupied LL, namely the ($n{-}1$)th LL, with $N{/}n$ orbitals in each LL. An example with $N{=}8$ and $n{=}2$ (i.e., $4$ electrons in each LL) is illustrated in Fig.~\ref{fig: 2_IQH_with_QH}, where different choices of the empty orbital in the SLL correspond to different $\phi_{2}^{N_{\rm QHs}=1,m}$. While Eq.~\eqref{eq: QH_coherent_state} is inspired by earlier QH composite fermion states~\cite{Gattu24}, we reformulate the construction, in particular to enforce the required antisymmetry, introducing $(-1)^{m-n+1}$, which is key to encoding non-Abelian properties. Specifically, this construction reproduces the bosonic Laughlin QH state up to some QH coordinate factors and is a general case of the bosonic Moore-Read state (detailed in the Supplemental Material (SM)~\cite{SM}). 

 We generalize this construction to $N_{\rm QHs}$ QHs localized at $\omega_1,\omega_2,\ldots,\omega_{N_{\rm QHs}}$ in an $n$-filled IQH factor, the corresponding wave function is given by:

\begin{equation}
\begin{aligned}
\Phi_n^{N_{\rm QHs}}
{=}
\sum_{m_1<\cdots<m_{N_{\rm QHs}}}
C_{m_1,\ldots,m_{N_{\rm QHs}}}
\!\left(\{ \omega \}\right)~
\Phi_n^{N_{\rm QHs},\,m_1,\ldots,m_{N_{\rm QHs}}}
\!\left(\{ z \}\right), 
\end{aligned}
\end{equation}
with 
\begin{equation}
C_{m_1,\ldots,m_{N_{\rm QHs}}}(\{\omega\})
=
\det\!\left[
(-1)^{m_j+n-1}\,
(\omega_a)^{m_j+n-1}
\right]_{a,j=1}^{N_{\rm QHs}},
\end{equation}
being a determinant of $N_{\rm QHs} {\times} N_{\rm QHs} $ matrix whose $(a,j)$ entry is the $m_j$-th LLL orbital evaluated at the $a$-th QH coordinate $\omega_a$.

\textbf{\textit{Parafermionic parton states.}}
We now focus on the $\Phi_2^k$ family, whose associated $SU(2)_k$ CFT sector can be realized by the product $\text{CFT}_{U(1)}\otimes \text{CFT}_{\psi_k}$, where $\text{CFT}_{\psi_k}$ is the $\mathcal{Z}_k$ parafermion CFT and $\text{CFT}_{U(1)}$ is a chiral boson CFT~\cite{Zamolodchikov85,Gepner87}. The chiral boson sector generates the Abelian $\Phi_1$ factors, implying that the non-Abelian CFT content of $\Phi_2^k$ is the same as that of the RR$k$ series. In particular, the $2^21$ state is Ising-like and therefore supports the same type of non-Abelian anyons as the MR state~\cite{Jain89b, Jain90}. Nevertheless, these phases are topologically distinct, as reflected in their Wen--Zee shift and chiral central charge~\cite{Wen92}.

Parton states of the form $\Phi_2\Phi_2\Phi_1^{p}$ have been discussed as candidate descriptions of even-denominator FQH plateaus in settings such as wide quantum wells, strong LL mixing, and higher LLs in graphene~\cite{Wu17, Faugno19, Kim19, Zhao23, Sharma22, Sharma23, Balram25, Bose25}. Several experimentally relevant plateaus may instead be described by the closely related states $[\Phi_2^k]^*\Phi_1^{k+1}$~\cite{Balram18, Balram19}, which carry the same non-Abelian content as $\Phi_2^k$ but with opposite edge chirality and are argued to lie in the universality class of the particle--hole conjugate of RR$k$. Moreover, extensive numerical work indicates that LL mixing can favor these states at the experimentally observed plateaus~\cite{Ambrumenil88, Balram13b, Morf98, Scarola02, Pakrouski15, Rezayi17, Wojs09, Zhu15, Mong15, Pakrouski16}. A practical advantage of the parton approach is that it provides trial wave functions in this universality class that remain numerically tractable at much larger system sizes. We therefore present results for the analytically simpler $\Phi_2^k$ series, with the same conclusions applying to $[\Phi_2^k]^*$. 
Unless stated otherwise, we work with unprojected parton states.

\textbf{\textit{Fusion Basis and Level-Rank Duality.}} In a parton state $\Phi_n^m$ with $N_{\rm QHs}{>}1$, QHs can be distributed among the $m$ factors in multiple ways. In particular, for the $\Phi_2^2$ state with four anyons [two QHs and two QPs], Ref.~\cite{Wen91} proposed that the degenerate Hilbert space is exhausted by two distributions of the QH-QP pairs among the two $\Phi_2$ factors:
(i) $\Phi_2^{(N_{\rm QHs}=2,N_{\rm QPs}=2)}\,\Phi_2$, and
(ii) $\Phi_2^{(N_{\rm QHs}=1,N_{\rm QPs}=1)}\,\Phi_2^{(N_{\rm QHs}=1,N_{\rm QPs}=1)}$.
While this distribution prescription produces the correct counting of independent states in this minimal case, it does not reproduce the non-Abelian braiding matrix of the four-QH MR~\cite{Nayak96}. The reason is that configuration (i) is effectively Abelian, since exchanges act as row permutations within a single determinant, the state maps back to itself under any exchange, up to an overall phase and sign. Therefore, it is crucial to identify the different possible distributions and, for each choice, determine the associated degenerate Hilbert space under particle exchanges, as detailed in the SM~\cite{SM}.

Here we focus on the choice in which the $N_{\rm QHs}$ QHs are equally divided among the $m$ factors (assuming $N_{\rm QHs}$ is divisible by $m$), corresponding to a distribution
$t{=}(t_1,t_2,{\ldots},t_m)$ with $t_i{=}N_{\rm QHs}{/}m$ [the number of QHs in the $i$th factor] for all $i$. This is one of many possible distributions, with the full set $D(N_{\rm QHs},m)$ (a few examples are given in SM~\cite{SM}). This choice is motivated by the analogy between $\Phi_2^{k}$ parton states and the bosonic RR$k$ states. The bosonic RR$k$ wave function is obtained by symmetrizing over partitions of $N$ particles into $k$ layers (with $N$ divisible by $k$), with the electrons in each layer forming $\nu{=}1{/}2$ bosonic Laughlin state~\cite{Cappelli01}. Excitations are then introduced by placing $N_{\rm QHs}/k$ QHs in each layer and symmetrizing over all indices, with a $\nu{=}1$ Laughlin form. In the parton language, the natural analogue is to place $N_{\rm QHs}/k$ QHs in each $\Phi_2$ factor (refer to SM for more details~\cite{SM}). For each distribution $t{\in} D(N_{\rm QHs},m)$, there is an associated Hilbert space $\mathcal H[t]$ that is closed under QH exchanges (analogous to the Nayak--Wilczek indexing~\cite{Nayak96}), spanned by the fusion-basis states $\Psi_{\nu}^{N_{\rm QHs},\,\mathcal{H}_j[t]}$. However, the resulting states are not all linearly independent. We numerically determine the number of independent states for various $n$ and $m$ across multiple system sizes, as summarized in Table~\ref{tab: no._of_indp_states}, where $O_{\rm num}$ is the numerically obtained count and $O_{\rm exp}$ is the CFT expectation from the fusion rules of the relevant primary operators. We observe that our construction yields the correct rank expected from the fusion rules of the underlying CFT (e.g., the $\mathcal{Z}_k$ parafermion CFT for the $\Phi_2^k$ family~\cite{slingerland01}). Furthermore, our approach yields rank predictions for higher-order parton states whose fusion-space dimensions have not yet been worked out (e.g., $\Phi_3^2$ and $\Phi_2^5$).

\setlength{\tabcolsep}{8pt} 
\renewcommand{\arraystretch}{1.1} 

\begin{table}[t!]
    \begin{center}
    \resizebox{\linewidth}{!}{%
        \begin{tabular}{ |c|c|c|c|c|c| } 
            \hline
            $N_{\rm QH}$ & \centering $t$ &  $\Phi_n^m$ & $O_{\rm num}$ & $O_{\rm exp}$ & $LRD$ \\
            \hline
            \multirow{2}{3em}{$4$} & \multirow{2}{3em}{$(2,2)$} & $\Phi_1^2$ & $1$ & $1$  & $\Phi_2$  \\ 
            & & $\Phi_2^2$ & $2$ & $2$ & $\Phi_2^2$   \\ 
            \hline
            \multirow{5}{3em}{$6$} & \multirow{3}{3em}{$(3,3)$} &
            $\Phi_1^2$ & $1$ & $1$ & $\Phi_2$ \\ 
            & & $\Phi_2^2$ & $4$ & $4$ & $\Phi_2^2$ \\ 
            & & $\Phi_3^2$ & $5$ & $5$ & $\Phi_2^3$  \\ \cline{2-6}
            & \multirow{2}{3em}{$(2,2,2)$} & $\Phi_1^3$ & $1$ & $1$ & $\Phi_3$  \\ 
            &   & $\Phi_2^3$ & $5$ & $5$ & $\Phi_3^2$  \\
            \hline
            \multirow{6}{3em}{$8$}& \multirow{4}{3em}{$(4,4)$} & $\Phi_1^2$ & $1$ & $1$ & $\Phi_2$  \\ 
            & & $\Phi_2^2$ & $8$ & $8$ & $\Phi_2^2$  \\ 
            & & $\Phi_3^2$ & $13$ & $13$ & $\Phi_2^3$ \\ 
            & & $\Phi_4^2$ & $14$ & $14$ & $\Phi_2^4$ \\ \cline{2-6} 
            & \multirow{2}{3em}{$(2,2,2,2)$}& $\Phi_1^4$ & $1$ & $1$ &  $\Phi_4$\\
            & & $\Phi_2^4$ & $14$ & $14$ &  $\Phi_4^2$\\
            \hline
            \multirow{3}{3em}{$9$} & \multirow{3}{3em}{$(3,3,3)$}& $\Phi_1^3$ & $1$ & $1$ & $\Phi_3$  \\ 
            & & $\Phi_2^3$ & $21$ & $21$ & $\Phi_3^2$  \\ 
            & & $\Phi_3^3$ & $42$ & $-$ & $\Phi_3^3$ \\ 
            \hline
            \multirow{7}{3em}{$10$} & \multirow{5}{3em}{$(5,5)$} & $\Phi_1^2$ & $1$ & $1$ & $\Phi_2$  \\ 
            & & $\Phi_2^2$ & $16$ & $16$ & $\Phi_2^2$  \\ 
            & & $\Phi_3^2$ & $34$ & $34$ & $\Phi_2^3$ \\ 
            & & $\Phi_4^2$ & $41$ & $41$ & $\Phi_2^4$ \\
            & & $\Phi_5^2$ & $42$ & $-$ & $\Phi_2^5$ \\ \cline{2-6}            
            &\multirow{2}{*}{$(2,2,2,2,2)$}& $\Phi_1^5$ & $1$ & $1$ & $\Phi_5$ \\
            & &$\Phi_2^5$ & $42$ & $-$ & $\Phi_5^2$ \\
            \hline
        \end{tabular}}
        
    \end{center}
    \caption{This table presents the numerically extracted rank $O_{\rm num}$ of the QH fusion-basis states in $\mathcal{H}[t]$ for a given distribution $t$ in the state $\Phi_n^m$ with $N_{\rm QHs}$ QHs,  alongside the theoretical expectation $O_{\rm exp}$~\cite{slingerland01}. For each $\Phi^n_m$, we also outline the level-rank dual (LRD) partner $\Phi_m^n$. }
    \label{tab: no._of_indp_states}
\end{table}

Beyond the numerical benchmarks, the robustness of these results can be understood from a structural analysis of the parton wave functions (see SM~\cite{SM}). In particular, $\Phi_2^2$ can be viewed as a natural generalization of the bosonic MR state, wherein rather than assigning exactly $N{/}2$ particles to each of two layers, the parton construction allows particles to populate both sectors, thereby generating additional Abelian contributions~\cite{SM}. This interpretation is consistent with the corresponding field-theory description~\cite{Henderson24} and extends straightforwardly to the $k{>}2$ parafermionic parton families. We further benchmark our rank results for the projected states $\mathcal{P}_{\rm LLL}\Phi_2^k\Phi_1^{2k}{\sim}\big(\Psi^{\rm Jain}_{2/5}\big)^k$, using the numerically tractable Jain--Kamilla projection~\cite{Jain97,Jain97b,Jain07}, which could underlie FQH states observed in wide quantum wells~\cite{Faugno19, Sharma23, Balram25}. Finally, leveraging the flexibility of the parton framework, we test the level--rank duality predicted by CS theory: a level-$n$ $SU(m)$ theory is dual to a level-$m$ $SU(n)$ theory, $SU(m)_n {\leftrightarrow} SU(n)_m$~\cite{Hsin16}. 
Since this duality equates the topological data (such as fusion rules and braiding), it predicts that the fusion-space dimensions for the $\Phi_2^{k}$ family (described by $SU(2)_k$ WZW) match those for $\Phi_k^{2}$ (described by $SU(k)_2$ WZW), as summarized in Table~\ref{tab: no._of_indp_states}.

\textbf{\textit{Braiding Matrix.}} The RR$k$ wave function can be formulated as the correlation function of the primary fields in the $\mathcal{Z}_k$ parafermion theory~\cite{Zamolodchikov85}. Distinct fusion channels (i.e., different ways the QH operators fuse to $\mathbf{1}$) correspond to different conformal blocks, which form the many-body QH wave functions and transform into one another under monodromies of the QH coordinates $\{\omega\}$, i.e., braiding operations. Similarly, the conformal block wave function in the $i$th fusion channel using parton states can be expressed as
\begin{equation} 
\Psi^{\text{CB},i}_{\nu}(\{ \omega \}, \{ z \})=\sum_{j=1}^{O_{\rm exp}} A_{ij}(\{ \omega \})\Psi_{\nu}^{N_{\rm QHs}, \,\mathcal{H}_j[t]} (\{ \omega \}, \{ z \}), 
\label{eq: conformal_block_parton} 
\end{equation} 
which is a superposition of the parton QH fusion basis states $\Psi_{\nu}^{N_{\rm QHs}, \, \mathcal{H}_j[t]}$, which resides in the space $\mathcal{H}[t]$ associated with the distribution configuration $t$, with coefficients $A_{ij}(\{\omega\})$. In conformal-block wave functions, exchanging QHs $a$ and $b$ ($a{\leftrightarrow} b$) acts through the multivalued analytic structure encoded in the coefficients $A_{ij}(\{\omega\})$ (via branch cuts) along with the fusion-basis states~\cite{Ardonne07}. This action is captured by the braiding matrix $\mathcal{B}^{a\leftrightarrow b}$, which can also be obtained purely from CFT data without constructing many-body wave function\cite{Moore89}. In the many-body approach, the braiding matrix $\mathcal{B}^{a{\leftrightarrow} b}$ can be obtained by orthonormalizing the conformal-block basis and computing overlaps, $\mathcal{B}^{a\leftrightarrow b}_{ij}={\langle }\Psi_{\nu,\mathrm{initial}}^{\mathrm{CB},j}{\mid }\Psi_{\nu,\mathrm{final}}^{\mathrm{CB},i}{\rangle}$, where $\Psi_{\nu,\mathrm{initial}}^{\mathrm{CB},j}$ is the $j$th conformal block at the initial QH configuration and $\Psi_{\nu,\mathrm{final}}^{\mathrm{CB},i}$ is the $i$th conformal block obtained after exchanging QHs $a$ and $b$ (see SM~\cite{SM}).

Using the method described above, we numerically compute the $N_{\rm QHs}{=}4$ braiding matrices for the $\Phi_2^2$ and $\Phi_2^3$ parton states, using the known coefficients $A_{ij}(\{\omega\})$ for the MR and RR$3$ four-QH conformal blocks (see SM~\cite{SM})~\cite{Ardonne07}. Since $A_{ij}(\{\omega\})$ is fixed entirely by correlators of the QH primary fields in the underlying CFT (e.g., the $\sigma$ field for the MR state~\cite{Nayak96}) and $\Phi_2^k$ generalizes the bosonic RR$k$ states (as motivated above)[note, the bosonic and fermionic RR$k$ wave functions differ only by Abelian $\Phi_1$ factors], these QH coefficients carry over for the corresponding parton states, up to additional factors (see SM~\cite{SM}), serving as a canonical basis to benchmark our many-body calculation. We choose $t{=}(2,2)$ [for $\Phi_2^3$, one of the $\Phi_2$ factor has no QHs], resulting in the subspace $\mathcal{H}(2,2){=}\{(1,2)(3,4),(1,3)(2,4),(1,4)(2,3)\}$, with a linear dependence among these three states that matches the known four-QH relation for the RR$k$ sequence~\cite{Ardonne07,SM}. 
\begin{figure}[htbp!]
\includegraphics[clip,width=1\columnwidth]{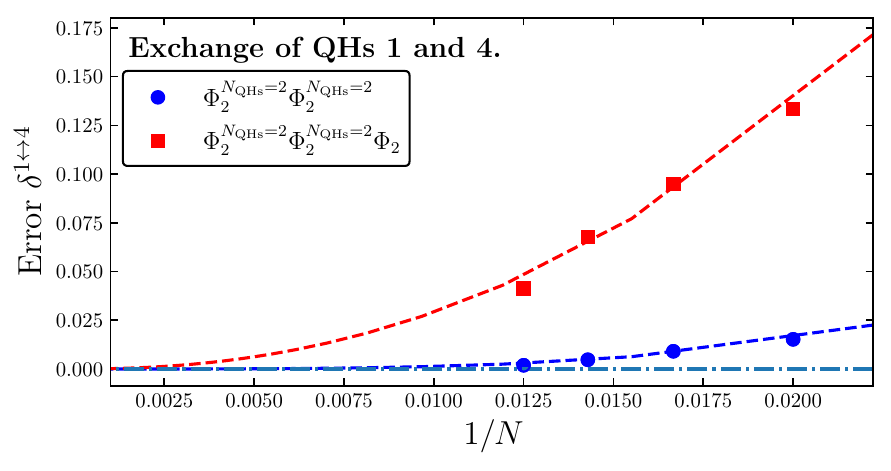} 
\caption{QHs $1$ and $4$ are initially placed at distance of $r{=}R{/}3$ from the origin at polar angles $\theta{=}0$ and $\pi$, respectively, where $R{=}\sqrt{2N{/}\nu}$. QHs $2$ and $3$ are fixed at  $r{=}2R{/}3$ at $\theta{=}0$ and $\pi$, respectively.
 }
\label{fig: Error_vs_N_inv}
\end{figure}

To benchmark our numerically extracted braiding matrix against the CFT prediction, we quantify their deviation for the exchange $a{\leftrightarrow}b$ by
$\delta^{a{\leftrightarrow}b}{=}\sqrt{\sum_{i,j}\big|\mathcal{B}^{a{\leftrightarrow}b}_{{\rm exp},ij}-\mathcal{B}^{a{\leftrightarrow}b}_{{\rm num},ij}\big|^{2}}$,
where $\mathcal{B}^{a{\leftrightarrow}b}_{{\rm exp}, ij}$ denotes the CFT value and $\mathcal{B}^{a{\leftrightarrow}b}_{{\rm num}, ij}$ our numerically obtained $(i,j)$ matrix element. Figure~\ref{fig: Error_vs_N_inv} shows $\delta^{1{\leftrightarrow}4}$ as a function of system size. We find that the extracted braiding matrices agree closely with the theoretical predictions for both MR and RR$3$, with the agreement improving as $N$ increases (as QHs become more widely separated and their overlaps decrease); (see SM for the other exchanges at $N{=}80$)~\cite{SM}. While MR braiding has been previously obtained via adiabatic simulations~\cite{Tserkovnyak03}, analogous calculations for RR$3$ have been hindered by the symmetrization bottleneck. Our approach instead extracts the braiding matrix from a single overlap between the initial and exchanged conformal-block states, avoiding the need to simulate an entire adiabatic trajectory (which accumulates additional Abelian Berry phase~\cite{Jeon04, Bose24b, Gattu24}). This enables a many-body extraction of Fibonacci-anyon braiding data up to $N\!\sim\!80$. Consistent with level--rank duality, we find that $\Phi_2^3$ and $\Phi_3^2$ have the same rank and identical braiding matrices for the four-QH case. We also obtain similar results for the projected $\Phi_2^2$ and $\Phi_2^3$ states, with projection leaving the anyonic data unchanged~\cite{Balram15a, Balram16b, Anand22}.

\textbf{\textit{Discussion.}} Our construction of parton QH basis states extends naturally to a larger number of QHs and to states whose braiding data have not yet been determined from CFT. While here we work with conformal-block wave functions, which require the QH coefficients (similar to past works~\cite{Tserkovnyak03}), it would be valuable to extract braiding data directly from the QH basis itself. The same framework should also extend to QP sectors, which are naturally accessible in parton constructions. Finally, recent experiments report signatures of non-Abelian anyons in even-denominator states~\cite{Kim26}, overcoming a long-standing complication of observing multiples of the fundamental charge due to anyon bunching~\cite{Biswas21, Ghosh25}. Given existing theoretical accounts in composite fermion settings~\cite{Gattu25}, our microscopic program may help interpret and guide analogous studies for parton phases. We leave these interesting directions for future work.

\textbf{\textit{Acknowledgments}.} I want to thank Ajit C. Balram for the useful discussions. Computational portions of this research were conducted using the Nandadevi supercomputer maintained and supported by the Institute of Mathematical Sciences' High-Performance Computing Center.

\bibliography{biblio_fqhe}
\end{document}


\title{Supplemental Material for ``Non-Abelian fusion and braiding in many-body parton states"}

\author{Koyena Bose}
\email{koyenab@imsc.res.in}
\affiliation{Institute of Mathematical Sciences, CIT Campus, Chennai 600113, India}
\affiliation{Homi Bhabha National Institute, Training School Complex, Anushaktinagar, Mumbai 400094, India} 

\maketitle

\setcounter{equation}{0}
\setcounter{figure}{0}
\setcounter{table}{0}
\setcounter{page}{1}
\setcounter{section}{0}

\makeatletter
\renewcommand{\theequation}{S\arabic{equation}}
\renewcommand{\thefigure}{S\arabic{figure}}
\renewcommand{\thesection}{S\Roman{section}}
\renewcommand{\thepage}{\arabic{page}}
\renewcommand{\thetable}{S\arabic{table}}

This supplemental material (SM) provides explanations of the results presented in the main text.


\section{Quasiholes in Parton States}
\label{sec: excited_parton_states}

For any unprojected parton state $\Psi_{\nu}{=}\prod_{\lambda} \Phi_{n_\lambda}$, the $n_\lambda{=}n$ integer quantum Hall (IQH) factor is constructed by occupying $N/n$ orbitals in each of the $n$ pseudo Landau levels, called Lambda levels ($\Lambda$Ls). In general, the single-particle state in the $m$th orbital of the $n$th LL is given by
 \begin{equation}
      \eta_{n, m}(z)= \frac{(-1)^n}{\sqrt{2\pi}} \sqrt{\frac{n!}{2^m(m+n)!}} \bigg (\frac{z}{\ell} \bigg)^m L_n^m \bigg( \frac{|z|^2}{2\ell^{2}}\bigg) e^{-\frac{|z|^2}{4\ell^{2}}},
    \label{eq: single-particle states}
 \end{equation}
where $n{=}0,1,\ldots$, $m{=}{-}n,{-}n{+}1,\ldots$, and $\ell{=}\sqrt{\hbar c{/}(eB)}$ is the magnetic length at the magnetic field $B$. Accordingly, we write $\Phi_n$ in a simplified form, omitting overall constants and terms equivalent under elementary row/column operations, which can be reduced to~\cite{Jain07}:
 \begin{equation}
    \Phi_{n} =\begin{vmatrix}
     . & .  & .& . & . \\  
	  . & .  & .& . & . \\  
    \bar{z}_1z_1^{N/n-1} & \bar{z}_2z_2^{N/n-1} & . & . & .\\ 
    . & . & . & . & . \\ 
	\bar{z}_1z_1 & \bar{z}_2z_2  & . & . & . \\ 
    \bar{z}_1 & \bar{z}_2  & . & . & . \\ 
	z_1^{N/n-1} & z_2^{N/n-1} & . & . & .\\ 
    . & . & . & . & .\\ 
    . & . & . & . & .\\
	z_1 & z_2 & . & . & . \\ 
    1 & 1 & . & . & . \\ 
	\end{vmatrix} \text{exp} \bigg ( -\sum_j |z_j|/4 \ell_{n}^{2} \bigg)
    \label{eq: phi_det}. 
\end{equation}
where $\ell_{n}$ is the effective magnetic length corresponding to the $n_\lambda{=}n$ factor, which obeys $\sum_{\lambda} 1/ \ell_{n_\lambda}^2=1/\ell^2$ (another version of filling fraction constraint, i.e, $1/\nu{=}\sum_\lambda 1/n_\lambda$). So in the parton state, the magnetic length present in the exponential term from each factor adds up to give $1{/}\ell^2$, while any $\ell_{n_{\lambda}}$ present in the single particle orbitals within the determinant can be absorbed into overall constants.

We now describe how to construct quasiholes (QHs) in a parton state. Creating a hole in a given IQH factor and then multiplying by the remaining factors effectively attaches flux to the hole so that it experiences a modified magnetic field and behaves as a QH. To set notation, we first create a hole in a single $\Phi_n$ factor by removing an electron from the topmost filled Landau level (LL). We denote the resulting state by $\phi_n^{N_{\rm QHs}=1,m}$, a Slater determinant of $N$ electrons filling the lowest $n$ LLs, except for the $L_z{=}m$ orbital in the $(n\!-\!1)$th LL (with $N/n$ orbitals filled in each LL). To localize the hole at position $\omega$, which will be crucial for braiding operations, we define:

\begin{eqnarray}
    \Phi^{N_{\rm QHs}{=}1}_{n}&=&\sum_{m=-(n-1)}^{\infty} (-1)^{m+(n-1)}  \omega^{m+(n-1)} \phi^{N_{\rm QHs}=1,m}_{n} (\{ z_{i} \}) \nonumber \\
\label{eq: phi_n_1QH_localized}
\end{eqnarray}
which is motivated by the coherent-state construction in Ref.~\cite{Gattu24} and uses LLL orbitals ($\eta_{0,m}$) for the QHs with overall constants suppressed. Note, for a finite system, the number of orbitals in the topmost filled LL is finite, and so is the sum over states $\phi_n^{N_{\rm QHs},m}$ for different values of $m$ at any $n$ and $N_{\rm QHs}$. This hole becomes a QH of the full parton state once it is multiplied by the remaining IQH factors; however, for ease of notation, we will refer to it as a QH even before including those factors. 

\subsection{$\Phi_1$ Factor}
We now examine several QH constructions for different values of $n$. We begin with $n{=}1$ and a single QH localized at $\omega$:
\begin{eqnarray}
    \Phi^{N_{\rm QHs}{=}1}_{n}&=&\sum_{m=0}^{\infty} (-1)^m  \omega^m \phi^{N_{\rm QHs}=1,m}_{n} \nonumber \\
    &=& \prod_{i}(z_{i}-\omega) \prod_{j<k} (z_{j}-z_{k})~\text{exp} (-\sum_j |z_j|^2/4\ell_1^2).
\label{eq: Laughlin_1QH_localized}
\end{eqnarray} 
We extend our analysis to the case of two QHs localized at $\omega_1$ and $\omega_2$, which yields the following expressions:
\begin{eqnarray}
    \Phi^{\rm N_{\rm QHs}{=}2}_{1}&=&\sum_{m_1,m_2=0}^{\infty} (-1)^{(m_1+m_2)} \omega_1^{m_1} \omega_2^{m_2} \phi^{ N_{\rm QHs}{=}2}_{1} (\{ z_{i} \}) \nonumber \\
    &=& (\omega_{1}-\omega_{2})\prod_{i}(z_{i}-\omega_1)(z_{i}-\omega_2) \prod_{j<k} (z_{j}-z_{k})~\text{exp} (-\sum_j |z_j|^2/4\ell_1^2), 
\label{eq: phi_1_2QH_localized}
\end{eqnarray}
We have also examined the three-hole case and find that the wave-function differs only by multiplicative factors involving the quasihole coordinates, such as $(\omega_{3}-\omega_{1})$ and $(\omega_{3}-\omega_{2})$, while the remaining dependence is Laughlin-like. While we have focused on creating QHs in a single $\Phi_n$ factor, for $N_{\rm QHs}{>}1$ they can be distributed among any of the $\Phi_\lambda$ factors, which will be detailed in the Sec.~\ref{subsec: distribution_of_QHs}.

\subsection{$\Phi_2$ Factor}
We now focus on the ground state $\Phi_2$, which can be decomposed as outlined in Ref.~\cite{Henderson24}:
\begin{eqnarray}
	\Phi_{2} 
    &=&\begin{vmatrix}
   z_{1}^{N_1-1}\bar{z}_{1} & z_{2}^{N_1-1}\bar{z}_{2} & . & . & .  \\
    . & . & . & . & .\\
    . & . & . & . & .\\
    z_{1}\bar{z}_{1} & z_{2}\bar{z}_{2} & . & . & .  \\
    \bar{z}_{1} & \bar{z}_{2} & . & . & . \\ 
    z_{1}^{N_1-1} & z_{2}^{N_1-1} & .& . & . \\  
    . & . & . & . & .\\
    z_{1} & z_{2}  & . & . & . \\ 
	1 & 1 & . & . & . \\ 
	\end{vmatrix} \text{exp} \bigg ( -\sum_j |z_j|/4 l_B^2 \bigg) \nonumber \\
    &=& \sum_{P \in P_{N,N_1}} \text{sgn} (P)  \begin{vmatrix}
    z_{P(1)}^{N_1-1} & z_{P(2)}^{N_1-1} & .& . & z_{P(N_1)}^{N_1-1} \\  
    . & . & . & . & .\\
    z_{P(1)} & z_{P(2)}  & . & . & z_{P(N_1)} \\ 
	1 & 1 & . & . & 1 \\ 
	\end{vmatrix} \nonumber \\
    && \times \begin{vmatrix}
    z_{P(N_1+1)}^{N_1-1}\bar{z}_{P(N_1+1)} & . & .& . & z_{P(N)}^{N_1-1} \bar{z}_{P(N)} \\  
    . & . & . & . & .\\
    z_{P(N_1+1)}\bar{z}_{P(N_1+1)} & .  & . & . & z_{P(N)} \bar{z}_{P(N)}\\ 
	\bar{z}_{P(N_1+1)} & . & . & . & \bar{z}_{P(N)}  \\ 
	\end{vmatrix} \text{exp} \bigg ( -\sum_j |z_j|/4 l_B^2 \bigg) . 
    \label{eq: phi_2_expanded}
\end{eqnarray}
where sgn($P$) occurs as a result of anti-symmetry in the problem.  The set $P_{N, N_1}$ comprises all distinct permutations $P$ that splits $N$ indices into two subsets, $P_1$ and $P_2$, of sizes $N_1$ and $N-N_1$, respectively (in this case $N_1=N/2$). For each distinct permutation $P$, $P(i)$ denotes the index of the $i$th particle, with its coordinate given by $z_{P(i)}$. Notably, each term of the determinant factorizes into a product of two Jastrow factors: one from the lowest Landau level (LLL) orbitals $\Phi_1(\{z\}_{P_1})$ and the other from the second Landau level (SLL) orbitals $\Phi_1(\{z\}_{P_2})$. 

Another convenient way to express Eq.~\eqref{eq: phi_2_expanded} is
\begin{equation}
\Phi_2
=
\mathcal S\!\left[
\text{sgn}(P)\,
\Phi_1(\{ z \}_{P_1})\,
\Phi_1(\{ z \}_{P_2})\,
\prod_{i\in P_2}\bar{z}_{i}
\right],
\label{eq: phi_2_2_expanded}
\end{equation}
where $\mathcal S$ denotes symmetrization over all bipartitions $(P_1,P_2)$ into two disjoint groups of distinct indices. Using this relation, we express
\begin{align}
\Phi_2^2
&=
\mathcal S\!\left[
\text{sgn}(P)\,
\Phi_1(\{ z \}_{P_1})\,
\Phi_1(\{ z \}_{P_2})\,
\prod_{i\in P_2}\bar{z}_{i}
\right]\,
\mathcal S\!\left[
\text{sgn}(K)\,
\Phi_1(\{ z \}_{K_1})\,
\Phi_1(\{ z \}_{K_2})\,
\prod_{i\in K_2}\bar{z}_{i}
\right].
\end{align}
Expanding this product yields three classes of terms:
(i) $P_1{=}K_1$,
(ii) $P_1{\cap} K_1{\neq} 0$ with $P_1{\neq} K_1$, and
(iii) $P_1\cap K_1=0$. For classes (i) and (iii), we obtain respectively
\begin{align}
\text{(i)}\;&\Rightarrow\;
\mathcal S\!\left[
\Phi_1^2(\{ z \}_{P_1})\,
\Phi_1^2(\{ z \}_{P_2})\,
\prod_{i\in P_2}\bar{z}_{i}^{\,2}
\right],\\[2pt]
\text{(iii)}\;&\Rightarrow\;
\mathcal S\!\left[
\Phi_1^2(\{ z \}_{P_1})\,
\Phi_1^2(\{ z \}_{P_2})\,
\prod_{i=1}^{N}\bar{z}_{i}
\right].
\end{align}
Specifically, class (iii) can be rewritten as $\prod_{i=0}^N \bar{z}_i \Psi_1^{\rm MR}$, since the $\nu{=}1$ bosonic Moore-Read (MR) state is
\begin{equation}
\Psi^{\rm MR}_{1}
=
\mathbb{S}\!\left[
\Phi^{2}_{1}(z_{1},\ldots,z_{N/2})\,
\Phi^{2}_{1}(z_{N/2+1},\ldots,z_{N})
\right].
\label{eq: MR_symm}
\end{equation}
Each term in class (i) carries a different subset of $\bar z_i^2$ factors, making it difficult to factor them out uniformly. So, we expect this class to contribute only additional Abelian factors, which will be made clear from the field-theory discussion below.

For class (ii), the indices common to both $P_1$ and $K_1$ contribute only Abelian correlation terms and carry no $\bar z_i$ dependence. These particles lie in the LLL sector of both $\Phi_2$ factors, so they generate Vandermonde correlations (factors of $z_i-z_j$) with the remaining indices in $P_1$ and $K_1$. Since the indices outside $P_1{\cap} K_1$ are disjoint between the two factors, the common particles correlate with all remaining LLL indices (across both the factors), yielding purely Abelian (Jastrow-type) terms. The same holds in the SLL sector: indices in $P_2{\cap} K_2$ again contribute only Abelian factors, now accompanied by $\bar z_i^{\,2}$. While the indices not in the intersections are distributed differently across the two $\Phi_2$ factors and therefore produce correlations specific to each factor. Moreover, because within each $\Phi_2$ the LLL and SLL sets are disjoint, an index missing from $P_1$ must appear in $K_2$, and an index missing from $K_1$ must appear in $P_2$. These complementary indices generate additional Vandermonde correlations, effectively promoting the corresponding Jastrow factors to second power. Summing all class (ii) contributions allows the Abelian terms to be extracted, leaving a bosonic MR-like structure acting on a reduced subset of indices (set by the number of disjoint indices) accompanied with $\bar{z}_i$ terms. Thus, $\Phi_2^2$ exhibits a $\nu{=}1$ MR-like structure dressed by additional Abelian correlation factors and $\bar z$ terms, in agreement with the expectations from field theory~\cite{Henderson24}:
\begin{eqnarray}
   \Psi_1 &=& \sum_{P,K \in P_{N,N_1}} \text{sgn(P)sgn(K)} \bra{0} C(N) [\prod_{i=P_1 \cup K_1}V_{2}(z_i)] \nonumber \\
    &&  [\prod_{k=P_2 \not\cap K_2} \bar{z}^2_k V_4(z_j)][\prod_{j=P_2 \cap K_2} \sqrt{2} \bar{z}_j V_3(z_j)] \ket{0} \nonumber \\
    &=& \bra{0} C(N) \prod_{i=1}^{N} (V_{2}(z_i) + \sqrt{2} \bar{z}_i V_3(z_i) + \bar{z}^2_i V_4(z_i)) \ket{0} \nonumber \\
     &=& \bra{0} C(N) \prod_{i=1}^{N} \Omega(z,\bar{z}) \ket{0} .
     \label{eq: operator_form_phi2_phi2}
\end{eqnarray}
In the expression above, $C(N)$ is the background charge operator ensuring that the correlator has a net zero charge, and $V_i$ are the fields generating the Laughlin factors. In \cite{Henderson24}, the authors show that a parton state $\Phi_n^m$ can be generated by the repeated application of the generating fields $V_l(z)$, with $l{=}m,m+1,....,nm$, associated with the chiral algebra $\mathcal{A}(n)_m$. Furthermore, they showed that the chiral algebra $\mathcal{A}(n)_m$ can also be represented using the $U(1) {\otimes} SU(n)_m$ WZW fields~\cite{Witten89}. Using $U(1) {\otimes} SU(n)_m {\equiv} U(1) {\otimes} U(1) {\otimes} \mathcal{Z}_k$, the fermion operators was redefined as $\Omega(z, \bar{z})=:e^{i\Phi(z)}:(:e^{-i\varphi(z)}:+\sqrt{2}\bar{z}\psi(z)+\bar{z}^2:e^{i\varphi(z)}:)$, where $\psi$ is a Majorana field from Ising CFT, and $\varphi$ and $\phi$ are fields of chiral boson theory. By comparing the operator construction in Eq.~\eqref{eq: operator_form_phi2_phi2} with the decomposition in Eq.~\eqref{eq: phi_2_2_expanded}, we identify the $V_{3}$ field, the one accompanied by $\bar z_i$, as the ingredient that generates the bosonic MR structure and thus underlies the emergence of non-Abelian behavior, consistent with our structural analysis. In contrast, the vortex operators $V_{2}$ (with no $\bar z_i$) and $V_{4}$ (with $\bar z_i^{\,2}$) contribute only $U(1)$ sectors, giving rise to Abelian Jastrow factors [as also evident from the decomposition (like class (i) which has no term with only single power of $\bar{z}$)]. Therefore, our expansion scheme is consistent with field-theoretic expectations.

Another way to interpret this decomposition, which provides useful intuition, is through the lens of distributing particles among ``layers" (note they are not the same as LLs). The bosonic MR state can be viewed as a two-layer construction in which each layer forms a $\Phi_1^{2}$ factor. One chooses any $N/2$ particle indices for the first layer, assigns the remaining $N/2$ to the second layer, and then symmetrizes over all distinct choices, as in Eq.~\eqref{eq: MR_symm}. Note, this is a special case in which every index belongs to exactly one of the two layers. In a more general setting, an index may effectively contribute to either layer or to both, and this possibility is naturally captured by the $\Phi_2^{2}$ parton state. Furthermore, class (iii) corresponds to the subcase in which the two partitions are completely disjoint, reproducing the full bosonic MR state (with some additional $\bar{z}$ terms). More generally, indices in the overlaps $P_1\cap K_1$ and $P_2\cap K_2$ can be viewed as being shared between the two layers. In contrast, the remaining (disjoint) indices remain unique to each layer, thereby yielding a bosonic MR-like structure on the disjoint subset, dressed by additional Abelian and $\bar z$ factors as discussed above. 

We now turn to the QH sector. Upon introducing a single hole in $\Phi_2$, the wave function takes the form:
\begin{eqnarray}
    \Phi^{ N_{\rm QHs}{=}1}_{2}&=&\sum_{m=0}^{\infty} (-1)^{m} \omega_1^{m} \phi^{ N_{\rm QHs}=1,m}_{n} (\{ z_{i} \}) \nonumber \\
    &=& \mathcal{S}[~\Phi_1(\{z\}_{P_1})~\Phi_1^{N_{\rm QHs}=1}(\{z\}P_2)\prod_{i=P_2}\bar{z}_{i}], 
\label{eq: phi_2_1QH_localized}
\end{eqnarray} 
It is straightforward to see that each filled LL contributes its own $\Phi_1$ factor, and introducing holes in $\Phi_2$ corresponds to removing orbitals specifically from the SLL. Consequently, the $\Phi_1$ factor associated with the SLL (i.e., the $P_2$ sector) acquires a Laughlin-like QH form, as in Eq.~\eqref{eq: Laughlin_1QH_localized}. Even upon multiplying an additional $\Phi_2$ into Eq.~\eqref{eq: phi_2_1QH_localized}, the QH coordinates reside entirely in this $\Phi_1$ factor. Similar to a single QH in the bosonic MR state, where the QH dependence is carried by a $\nu{=}1$ Laughlin (i.e., a $\Phi_1$ factor), we have $\Psi^{{\rm MR}, N_{\rm QHs}{=}1}_{1}{=}\mathbb{S}[\Phi^{N_{\rm QHs}=1}_{1}(z_{1},\ldots,z_{N/2})\Phi_{1}(z_{1},\ldots,z_{N/2})\Phi^{2}_{1}(z_{N/2+1},\ldots,z_{N})]$. For multiple QHs, there are different ways to distribute them between the two $\Phi_2$ factors. In particular, for the $k$-clustered Read–Rezayi states, the QHs are distributed equally among $k$ layers. In the parton language, this corresponds to distributing them equally among the $k\Phi_2$ factors, i.e., placing QHs in the distinct $\Phi_1$ factors associated with the SLL sector of each $\Phi_2$ factor. 

Similar grouping arguments can extend to higher-order parafermionic parton states, i.e., $\Phi_2^k$ with $k{>}2$, and lead to distinct classes analogous to those identified for $k{=}2$. The corresponding field-theory construction has been developed for higher $k$ [generalizing Eq.~\eqref{eq: operator_form_phi2_phi2}], and we expect it to be consistent with a direct structural analysis of the many-body wave functions. We leave a detailed treatment of this generalization and showing level-rank duality from structural analysis as future work.


\subsection{Distribution of QHs}
\label{subsec: distribution_of_QHs}

In this paper, we focus on the case $n_{\lambda}{=}n$ for all $\lambda$, i.e., parton states built from repeated copies of the same IQH factor. So, for $N_{\rm QHs}{>}1$, the QHs can be distributed among the $m$ copies in multiple ways. We generalize the mechanism as follows: we define $D(N_{\rm QHs},m)$ as the set of all possible distributions of $N_{\rm QHs}$ QHs in $m$ parton factors. Each configuration in this set is represented by a tuple $t{=}(t_1,t_2,\cdots,t_m)$, where $t_i$ denotes the number of QHs in the $i$th IQH factor, and $\sum_{i=1}^m t_i{=}N_{\rm QHs}$. Below, we have laid out a few examples for reference: 
\begin{enumerate}
    \item  $D(4,2){=}\{(0,4),(1,3),(2,2)\}$ 
    \item $D(6,2){=}\{(0,6),(1,5),(2,4),(3,3)\}$ 
    \item $D(6,3){=}\{C(6,2),(1,1,4),(1,2,3),(2,2,2)\}$,
\end{enumerate}

Each choice of distribution
$t{\in}D(N_{\rm QHs},m)$ defines a distinct subspace $\mathcal H[t]$~\cite{Nayak96}, spanned by a set of basis states labeled by $m$ \emph{micro-groups}. Concretely, a basis label consists of an $m$-tuple of micro-groups, $(g_1)(g_2)\cdots(g_m)$, where the $i$th micro-group $g_i$ contains exactly $t_i$ quasihole indices, as dictated by $t{=}(t_1,\ldots,t_m)$. The full subspace $\mathcal H[t]$ is generated by enumerating all distinct assignments of the $N_{\rm QHs}$ labels into micro-groups consistent with $t$ (modulo permutations within each micro-group). As an illustration, for $N_{\rm QHs}{=}4$ and $m{=}2$, the distribution $t{=}(2,2)$ gives the subspace $\mathcal H[(2,2)]$ spanned by the three micro-groupings
$(1,2)(3,4)$, $(1,3)(2,4)$, and $(1,4)(2,3)$. For example, the state labeled by $(1,2)(3,4)$ corresponds to the wave function
$\Psi_{\nu}^{N_{\rm QHs}=4,[(1,2)(3,4)]}$, where the micro-groups specify both how many QHs are assigned to each $\Phi_n$ factor and which QH labels they carry. This extends the micro-grouping construction of Nayak and Wilczek~\cite{Nayak96} to arbitrary parton states, allowing for all possible QH distributions. We list further examples below:

I. Basis state for each clustering corresponding to $C(4,2)$. 
\begin{enumerate}
    \item $\mathcal{H}[(0,4)]=\{(1,2,3,4) \}$
    \item $\mathcal{H}[(3)]=\{(1,2,3)(4),(1,2,4)(3),(2,3,4)(1)\}$
    \item $\mathcal{H}[(2,2)]=\{(1,2)(3,4),(1,3)(2,4),(1,4)(2,3)\}$
\end{enumerate}

II.  Basis state for each clustering corresponding to $C(6,2)$. 
\begin{enumerate}
    \item $\mathcal{H}[(0,6)]=\{(1,2,3,4,5,6)\}$
    \item $\mathcal{H}[(1,5)]=\{(1,2,3,4,5)(6),(1,2,3,4,6)(5),\cdots \}$
    \item $\mathcal{H}[(2,6)]=\{(1,2,3,4)(5,6),(1,2,3,5)(4,6),\cdots \}$
    \item $\mathcal{H}[(3,6)]=\{(1,2,3)(4,5,6), (1,2,4)(3,5,6),\cdots\}$
\end{enumerate}

III.  Basis state for each clustering corresponding to $C(6,3)$. 
\begin{enumerate}
    \item $\mathcal{H}[(1,1,4)]=\{(1)(2)(3,4,5,6),(1)(3)(2,4,5,6),\cdots\}$
    \item $\mathcal{H}[(1,2,3)]=\{(1)(2,3)(4,5,6),(1)(3,4)(2,5,6),\cdots\}$
    \item $\mathcal{H}[(2,2,2)]=\{(1,2)(3,4)(5,6),(1,2)(3,5)(4,6),\cdots\}$.
\end{enumerate}

\section{Conformal-Block Wave Functions}
The conformal block wave function for any $i$th fusion channel is given by 
\begin{equation} 
\Psi^{\text{CB},i}_{\nu}(\{ \omega \}, \{ z \})=\sum_{j=1}^{O_{\rm exp}} A_{ij}(\{ \omega \})\Psi_{\nu}^{N_{\rm QHs}, \,\mathcal{H}_j[t]} (\{ \omega \}, \{ z \}). 
\label{eq: conformal_block_parton} 
\end{equation} 
We will now focus on the case for $N_{\rm QHs}{=}4$ with $t{=}(2,2)$ for $\Phi_2^2$ and $\Phi_2^3$ (for $\Phi_2^3$, the distributions is $\Phi_2^{N_{\rm QHs}{=}2}\Phi_2^{N_{\rm QHs}{=}2}\Phi_2$). The fusion-space dimension is $2$ for both cases. Therefore, the co-efficients for $\Phi_2^2$ state is:
\begin{eqnarray}
    &1.& A_{00}=\frac{0.5(\omega_{12}\omega_{34})^{1/4}x^{1/4}\bigg(\sqrt{1-\sqrt{x}}+\sqrt{1+\sqrt{x}}\bigg )}{\omega_{12}\omega_{34}}\nonumber\\
    &2.& A_{01}=\frac{-0.5(\omega_{12}\omega_{34})^{1/4}x^{-1/4}(1-x)^{1/2}\bigg(-\sqrt{1-\sqrt{x}}+\sqrt{1+\sqrt{x}}\bigg )}{\omega_{13}\omega_{24}}\nonumber\\
    &3.& A_{10}=\frac{0.5(-1)^{-1/2}(\omega_{12}\omega_{34})^{1/4}x^{1/4}\bigg(-\sqrt{1-\sqrt{x}}+\sqrt{1+\sqrt{x}}\bigg )}{\omega_{12}\omega_{34}}\nonumber\\
    &4.& A_{11}=\frac{-0.5(-1)^{-1/2}(\omega_{12}\omega_{34})^{1/4}x^{-1/4}(1-x)^{1/2}\bigg(-\sqrt{1-\sqrt{x}}-\sqrt{1+\sqrt{x}}\bigg )}{\omega_{13}\omega_{24}}
\end{eqnarray}
with $w_{ij}{=}w_{i}-w_{j}$, $x{=}(\omega_{12} \omega_{34}){/}(\omega_{14}\omega_{32})$, $\mathcal{H}_0[(2,2)]=(12)(34)$, and $\mathcal{H}_1[(2,2)]=(13)(24)$. The coefficients are taken from Ref.~\cite{Ardonne07} (for the MR state), with additional $w_{ij}$-type factors appearing in the denominator. This adjustment is required because, as seen from Eq.~\eqref{eq: phi_1_2QH_localized}, different QH distributions among the IQH factors can introduce additional QH coordinate-dependent factors such as $w_{ij}$. Similarly, for the $\Phi_2^3$ state, we use the same fusion-basis states $\mathcal{H}_0[(2,2)]$ and $\mathcal{H}_1[(2,2)]$, with coefficients given by~\cite{Ardonne07}
\begin{eqnarray}
    &1.& A_{00}=\frac{(\omega_{12}\omega_{34})^{7/10}(1-x)^{1/10}{}_2F_1
(1/5,-1/5,3/5;x)}{\omega_{12}\omega_{34}}\nonumber\\
    &2.& A_{01}=\frac{-(\omega_{12}\omega_{34})^{7/10}(1-x)^{11/10}{}_2F_1
(6/5,4/5,8/5;x)}{3\omega_{13}\omega_{24}}\nonumber\\
    &3.& A_{10}=\frac{-(-1)^{2/5}C(\omega_{12}\omega_{34})^{7/10}x^{2/5}(1-x)^{1/10}{}_2F_1(1/5,3/5;7/5;x)
}{\omega_{12}\omega_{34}}\nonumber\\
    &4.& A_{11}=\frac{-2(-1)^{2/5}C(\omega_{12}\omega_{34})^{7/10}x^{-3/5}(1-x)^{11/10}{}_2F_1(1/5,3/5,2/5,x)}{\omega_{13}\omega_{24}}.
\end{eqnarray}
where ${}_2F_1(a,b;c;z)$ is the Gauss hypergeometric function and 
$C=\frac{1}{2}\sqrt{\frac{\Gamma(1/5)\,\Gamma^3(3/5)}{\Gamma(4/5)\,\Gamma^3(2/5)}}$, with $\Gamma(x)$ as the Gamma function.

\begin{table}[t!]
\centering
\small
\setlength{\tabcolsep}{6pt}
\renewcommand{\arraystretch}{1.25}
\begin{tabular}{|c|>{\centering\arraybackslash}m{0.40\linewidth}|>{\centering\arraybackslash}m{0.40\linewidth}|}
\hline
\multirow{2}{*}{Exchange}
& \multicolumn{2}{c|}{Braiding Matrix ($\Phi_2^{N_{\rm QHs}{=}2}\Phi_2^{N_{\rm QHs}{=}2}$)} \\
\cline{2-3}
& Exact & Numerical \\
\hline
$1 \leftrightarrow 2$/$3 \leftrightarrow 4$ &
$\begin{pmatrix}
1/\sqrt{2}(\equiv0.707107) - i/\sqrt{2}  & 0 \\
0 & 1/\sqrt{2} + i/\sqrt{2} 
\end{pmatrix}$ &
$\begin{pmatrix}
1/\sqrt{2} - i/\sqrt{2}  & 0 \\
0 & 1/\sqrt{2} + i/\sqrt{2} 
\end{pmatrix}$ accurate to $10^{{-}13}$ \\
\hline
$1 \leftrightarrow 3$/$2 \leftrightarrow 4$ &
$\begin{pmatrix}
1/\sqrt{2} & -1/\sqrt{2} \\
1/\sqrt{2} & 1/\sqrt{2}
\end{pmatrix}$ &
$\begin{pmatrix}
0.728(2)-0.002(2) i & -0.685(2) \\
0.685(2) & 0.728(2)+0.002(2) i
\end{pmatrix}$ \\
\hline
$1 \leftrightarrow 4$/$2 \leftrightarrow 3$ &
$\begin{pmatrix}
-i/\sqrt{2}  & -1/\sqrt{2} \\
1/\sqrt{2} & i/\sqrt{2}
\end{pmatrix}$ &
$\begin{pmatrix}
-0.002(2)-0.728(2) i & -0.685(2) \\
0.685(2) & -0.002(2)+0.728(2) i
\end{pmatrix}$ \\
\hline
\end{tabular}
\caption{Braiding matrices for different exchanges in the $\Psi_1^{\rm CB}$ (using $\Phi_2^2$) state with $N_{\rm QHs}{=}4$, with two QHs in each $\Phi_2$ factor. The numbers in parentheses indicate the statistical uncertainty from the Monte Carlo sampling.
}
\label{tab:braid_phi2sq}
\end{table}
\begin{table}[t!]
\centering
\tiny
\setlength{\tabcolsep}{6pt}
\renewcommand{\arraystretch}{1.25}
\resizebox{1.00\linewidth}{!}{%
\begin{tabular}{|c|>{\centering\arraybackslash}m{0.40\linewidth}|>{\centering\arraybackslash}m{0.40\linewidth}|}
\hline
\multirow{2}{*}{Exchange}
& \multicolumn{2}{c|}{Braiding Matrix ($\Phi_2^{N_{\rm QHs}{=}2}\Phi_2^{N_{\rm QHs}{=}2}\Phi_2$)} \\
\cline{2-3}
& Exact & Numerical \\
\hline
$1 \leftrightarrow 2$/$3 \leftrightarrow 4$ &
$\begin{pmatrix}
\cos(-3\pi/10)(\equiv0.587785) - i \sin(3\pi/10)(\equiv0.809017i) & 0 \\
0 & e^{+3\pi i/10}
\end{pmatrix}$ &
$
\begin{pmatrix}
0.58778526(5) - 0.80901698(9) i & 0 \\
0 & 0.58778526(9) + 0.80901698(5) i
\end{pmatrix}
$ \\
\hline
$1 \leftrightarrow 3$/$2 \leftrightarrow 4$ &
$\begin{pmatrix}
-e^{-9\pi i/10}\tau\equiv0.587785 + 0.190983 i & e^{-9\pi i/10}\sqrt{\tau} \\
-e^{9\pi i/10}\sqrt{\tau} & -e^{9\pi i/10}\tau
\end{pmatrix}$ &
$\begin{pmatrix}
0.609(2) + 0.193(2) i & -0.731(2)-0.237(2) i \\
0.731(2)-0.237(2) i & 0.609(2) - 0.193(2) i
\end{pmatrix}$ \\
\hline
$1 \leftrightarrow 4$/$2 \leftrightarrow 3$ &
$\begin{pmatrix}
-i\tau & -e^{\pi i/10}\sqrt{\tau} \\
-e^{-\pi i/10}\sqrt{\tau} & i\tau
\end{pmatrix}$ &
$\begin{pmatrix}
-0.003(3)-0.638(3) i & -0.731(2)-0.237(2) i \\
0.731(2)-0.237(2) i & -0.003(3)+0.638(3) i
\end{pmatrix}$ \\
\hline
\end{tabular}
}
\caption{Braiding matrices for different exchanges in the $\Psi_{2/3}^{\rm CB}$ (using $\Phi_2^3$) state with $N_{\rm QHs}{=}4$, with two QHs in each of two $\Phi_2$ factors and the third $\Phi_2$ factor with no QHs. The numbers in parentheses indicate the statistical uncertainty from the Monte Carlo sampling. [Note, $\tau{=}(\sqrt{5}{-}1)/2$.]}
\label{tab:braid_phi2cu}
\end{table}
\subsection{Braiding Matix}
The braiding matrix is a key topological datum, as it encodes how exchange of anyons transforms one degenerate state into a linear combination of the others, in a manner robust to local perturbations. In a conformal-block basis, these matrices are known for several CFTs (e.g., the $\mathcal{Z}_k$ parafermion theories). We start from the conformal-block basis $\{\,\ket{\Psi_{\nu,\mathrm{initial}}^{\mathrm{CB},i}}\,\}$ at a fixed quasihole configuration. After exchanging quasiholes $a{\leftrightarrow} b$ along a specified path, these evolve to $\{\,\ket{\Psi_{\nu,\mathrm{final}}^{\mathrm{CB},i}}\,\}$, related by the braiding matrix $B^{a{\leftrightarrow}b}$:
\begin{eqnarray}
\begin{pmatrix}
\ket{\Psi_{\nu,\mathrm{final}}^{\mathrm{CB},0}}\\
\ket{\Psi_{\nu,\mathrm{final}}^{\mathrm{CB},1}}
\end{pmatrix}
=
\begin{pmatrix}
B^{a{\leftrightarrow}b}_{00} & B^{a{\leftrightarrow}b}_{01}\\
B^{a{\leftrightarrow}b}_{10} & B^{a{\leftrightarrow}b}_{11}
\end{pmatrix}
\begin{pmatrix}
\ket{\Psi_{\nu,\mathrm{initial}}^{\mathrm{CB},0}}\\
\ket{\Psi_{\nu,\mathrm{initial}}^{\mathrm{CB},1}}
\end{pmatrix}.
\label{eq:braiding_def}
\end{eqnarray}
Equivalently,
\begin{equation}
\ket{\Psi_{\nu,\mathrm{final}}^{\mathrm{CB},i}}
=\sum_{j} B_{ij}\,\ket{\Psi_{\nu,\mathrm{initial}}^{\mathrm{CB},j}}.
\label{eq:braiding_expand}
\end{equation}
Taking overlaps with the initial basis gives
\begin{equation}
\Big\langle \Psi_{\nu,\mathrm{initial}}^{\mathrm{CB},k}\,\Big|\,\Psi_{\nu,\mathrm{final}}^{\mathrm{CB},i}\Big\rangle
=\sum_{j} B^{a{\leftrightarrow}b}_{ij}\,
\Big\langle \Psi_{\nu,\mathrm{initial}}^{\mathrm{CB},k}\,\Big|\,\Psi_{\nu,\mathrm{initial}}^{\mathrm{CB},j}\Big\rangle.
\label{eq:braiding_overlap}
\end{equation}
Now, if we take orthonormalize the intital basis, such that $\Big\langle \Psi_{\nu,\mathrm{initial}}^{\mathrm{CB},k}\,\Big|\,\Psi_{\nu,\mathrm{initial}}^{\mathrm{CB},j}\Big\rangle{=} \delta_{kj}$, we can extract any element of the braiding matrix as:
\begin{equation}
    B^{a{\leftrightarrow}b}_{ij}=\Big\langle \Psi_{\nu,\mathrm{initial}}^{\mathrm{CB},j}\,\Big|\,\Psi_{\nu,\mathrm{final}}^{\mathrm{CB},i}\Big\rangle.
    \label{eq:braiding_matrix_element}
\end{equation}

Using Eq.~\eqref{eq:braiding_matrix_element}, we extract the individual elements of the braiding matrix elements $B^{a{\leftrightarrow}b}_{ij}$ by evaluating the multi-dimensional integrals over the electron coordinates, which are carried out using the Metropolis Monte Carlo algorithm~\cite{Binder10}. We carry this out for several exchanges $a{\leftrightarrow}b$ using the $N_{\rm QHs}{=}4$ conformal-block states $\Psi_1^{{\rm CB},i}$ (corresponding to $\Phi_2^2$) and $\Psi_{2/3}^{{\rm CB},i}$ (corresponding to $\Phi_2^3$), restricting to the sector $t{=}(2,2)$ with fusion-space dimension $2$, and benchmark the results against the known CFT predictions. Table~\ref{tab:braid_phi2sq} and ~\ref{tab:braid_phi2cu} compare the CFT predictions with our numerical values obtained at the largest system size studied ($N{=}80$). Note that the exact braiding matrices reported in the table may differ from common conventions in the literature. We fix the overall $U(1)$ phase by dividing each element of the matrix by $\sqrt{\det [\mathcal{B}^{a{\leftrightarrow}b}]}$, thereby removing any associated Abelian phase. In addition, depending on the chosen exchange path (clockwise versus counterclockwise), our matrices may correspond to the inverse of those quoted elsewhere~\cite{Ardonne07}.

\bibliography{biblio_fqhe}